# Convolutional Long Short-Term Memory (convLSTM) for Spatio-Temporal Forecastings of Saturations and Pressure in the SACROC Field

Palash Panja[a,b,c,*], Wei Jia [a,c,d], Alec Nelson[a,d], Brian McPherson[a,c,d]

[a] National Energy Technology Laboratory, 626 Cochrans Mill Road, Pittsburgh, PA 15236, USA

[b] Department of Chemical Engineering, University of Utah, 50 S, Central Campus Dr., Room 3290 MEB, Salt Lake City, Utah 84112, USA

[c] Energy & Geoscience Institute, University of Utah, 423 Wakara Way, Suite 300, Salt Lake City, Utah 84108, USA

[d] Department of Civil and Environmental Engineering, University of Utah, 110 S, Central Campus Dr., Suite 2000, Salt Lake City, Utah 84112, USA

*Corresponding Author, *ppanja@egi.utah.edu*

## Abstract

A machine learning architecture composed of convolutional long short-term memory (convLSTM) is developed to predict spatio-temporal parameters in the SACROC oil field, Texas, USA. The spatial parameters are recorded at the end of each month for 30 years (360 months), approximately 83% (300 months) of which is used for training and the rest 17% (60 months) is kept for testing. The samples for the convLSTM models are prepared by choosing ten consecutive frames as input and ten consecutive frames shifted forward by one frame as output. Individual models are trained for oil, gas, and water saturations, and pressure using the Nesterov accelerated adaptive moment estimation (Nadam) optimization algorithm. A workflow is provided to comprehend the entire process of data extraction, preprocessing, sample preparation, training, testing of machine learning models, and error analysis. Overall, the convLSTM for spatio-temporal prediction shows promising results in predicting spatio-temporal parameters in porous media.

## Introduction

A hydrocarbon reservoir can be characterized by usually static properties such as permeability and porosity; and dynamic properties such as pressure and oil, gas, and water saturations. Forecasting these dynamic properties is important for economic production, maximum recovery, and future development. A reservoir like the Scurry Area Canyon Reef Operators Committee (SACROC) oilfield in Texas, USA, requires advanced techniques to forecast these spatial properties due to its geologic and operational complexities. There are several producers and injectors in this field for enhanced oil recovery (EOR). Water and $CO_2$ are injected alternatingly to improve the oil recovery from the production wells. The nature of this operation creates chaotic distributions of oil, water, and CO2 in the subsurface. As a consequence, the saturations and pressure maps exhibit complex patterns. In addition, the dynamic changes are also challenging to forecast. The reservoir simulator is the standard tool in the industry to predict production and to investigate the fluid and heat flows in the subsurface. Many advanced forecast techniques are being developed to avoid some limitations of reservoir simulators, such as model development requires time, expert supervision and robust data set. Moreover, computational time, memory, and power usage during simulation are also relatively higher compared to surrogate models and machine learning models.

Recent progress in machine learning algorithms employing deep neural networks encouraged researchers to model complex physical systems. Particularly, spatio-temporal machine learning algorithms are useful to address these types of issues. Several such machine learning algorithms have been developed recently for the prediction of various phenomena such as rainfall [1], flood [2], video frame [3-7], human movement to next location [8], sea surface temperature[9], stress field [10], surveillance event detection [11]. These models consider both temporal and spatial contexts. The majority of these algorithms is the variant of recurrent neural networks (RNN) and long short term memory (LSTM)[12] with the spatial context; for example, spatial temporal recurrent neural networks (ST-RNN)[8], predictive recurrent neural network (PredRNN and PredRNN++ )[3, 5], convolutional long short-term memory (convLSTM) [1, 4, 9, 11], spatio-temporal attention long short term memory model (STA-LSTM)[2], flexible spatio-temporal network (FSTN)[7], CostNet[6], physics-informed spatio-temporal long short term memory (Physics-Informed ST-LSTM)[10].

In this study, the convLSTM model is adopted to predict the dynamic saturation and pressure distributions at the SACROC unit. Shi et al.[1] first proposed convolutional long short term memory (convLSTM) by adding the spatial correlations using convolutional operation in long short term memory (LSTM) [12] to predict rainfall. The convLSTM is widely applied for video frame prediction from previous frames [3-7]. Other applications include but not limited to human gesture recognition [13], tongue motion[14], vegetation index[15], pedestrian attribute recognition[16], Arctic sea ice concentration[17], spatial-spectral feature extraction[18], scene text recognition [19], sea surface temperature (SST)[20], transportation mode [21]. There are not many applications of convLSTM in subsurface multiphase flow. For example, $CO_2$ leakage detection from a $CO_2$-EOR site[22, 23] was studied using bottom-hole pressure (BHP) data with convLSTM. A hybrid model was used to develop a numerical simulator with physical constraints for flow and transport in the subsurface environment [24]. Another hybrid model of convLSTM was successfully developed to predict the evolution of the $CO_2$ saturation and pressure in porous media during sequestration [25, 26].

**Reservoir Model**

The results from a simulated reservoir model representing a real field are extracted and analyzed for spatio-temporal data for this study. The reservoir model represents the Scurry Area Canyon Reef Operators Committee (SACROC) oilfield located on the eastern edge of the Permian Basin, Texas, USA. Details of this oilfield and the reservoir model are described in our previous study [27]. The SACROC reservoir model with 13,600 totals cells ($34 \times 16 \times 25$) is shown in Figure 1. The model has 23 producers and 22 injectors with the perforations in layers 19, 20, 21, and 22. The wells are placed based on a 5-spot injection pattern. The injection wells are alternated between $CO_2$ and water injections on an annual basis. Some producers are shut in randomly after a few years of production to mimic practical field scenarios. This may be caused by many factors like operational constraint in the downstream plants, supply-demand balance, oil price drops, maintenance issues, wellbore damage, formation damage, etc.

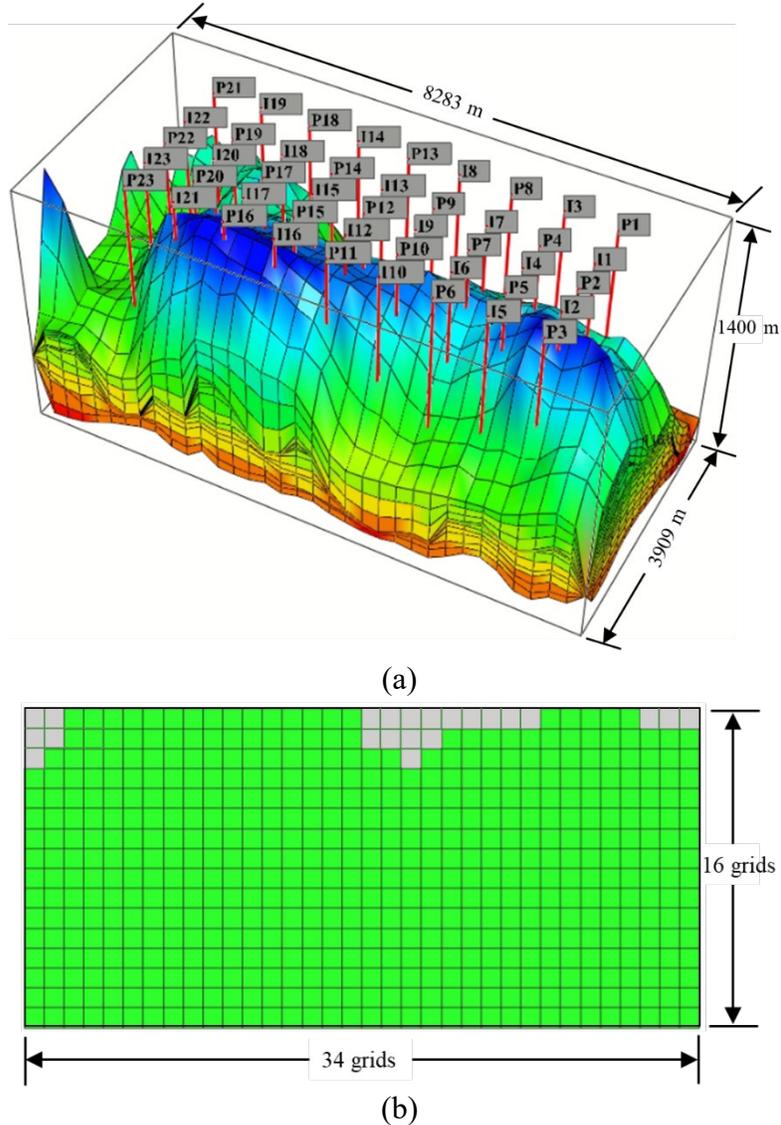

(a)

(b)

Figure 1: (a) Three-dimensional SACROC reservoir model[27] (b) horizontal layer 21 showing the active (green) and inactive (gray) cells

The model is built and simulated using a commercial simulator CMG-GEM package [28]. The model is run for 30 years duration. The reservoir model can be accessed at the National Energy Technology Laboratory's (NETL) repository Energy Data Exchange (EDX)[29].

In this study, a single horizontal layer (layer 21) that has the most well perforation is selected to evaluate the applicability of convLSTM in the SACROC unit. Therefore, this layer can be considered as a 2-dimensional data frame of size 34 × 16. Data are recorded monthly for the 30 years of operation, and as a consequence, a total of 360 of the 2-dimensional frame for each property (oil, gas, and water saturations, and pressure) is extracted.

It can be noted that although the numbers of cells in the x- and y- directions are 34 and 16, respectively, some cells are not active on this layer of interest (layer 21) as shown in Figure 1b. The objective of spatiotemporal machine learning is to predict a number of temporal sequences of tensors in the future provided the series of previous sequences.

**Workflow**

A complete workflow describing the entire process of data extraction, sample preparation, training and testing of the ML model, and performance analysis is shown in Figure 2.

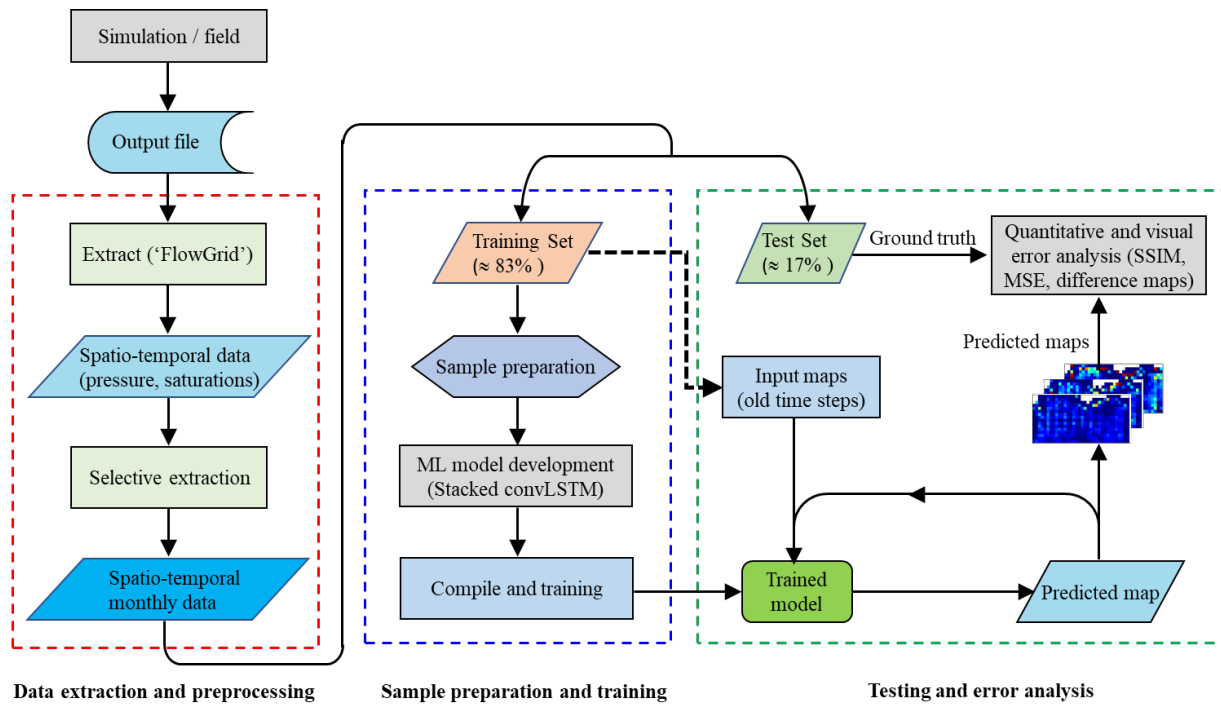

Figure 2: workflow describing the entire process of data extraction, sample preparation, the training of ML model, testing, and performance analysis

The workflow is broadly divided into three parts: data extraction and preprocessing, sample preparation and training of ML algorithm, and testing and error analysis. The connectivity of various Python scripts and information flow are shown in Appendix C. The first few steps for data extraction involve reading the relevant properties in output files generated by the reservoir simulator. An in-house Python script (ReGrid) is written to execute the job automatically[30, 31]. The script gathers spatio-temporal information such as pressure and saturations at all active grids for all time steps with a certain interval. The time interval is defined by the modeler during building the input file for the simulation. A file for a single time and each property at all cells is saved in the .npz file format. For this machine learning application, the data of the horizontal layer 21 at the end of each month for 360 months or 30 years is assembled from those .npz files.

The properties in general are normalized between 0 to 1 (or -1 to 1) in the majority of data analytic and machine learning tools. There are a few reasons for the normalization, such as transforming different variables into the same range to apply them in a unified model structure or equation. Another reason is that the output of some activation functions in machine learning is bound between 0 to 1, e.g., sigmoid, or -1 to 1, e.g., hyperbolic tangent. Therefore, the outputs (a few inputs are the same as outputs in the convLSTM or LSTM model) must be converted between the ranges of activation function that is used in the machine learning algorithm. The sigmoid activation function is used in this study, therefore, absolute values of pressures are transformed between 0 to 1. All other properties such as oil, water, and gas saturations inherently have values between 0 to 1, therefore normalization is not required for these properties.

The normalized frames are then divided into training and test sets. The input samples to the ML algorithm are prepared from the training set, which is discussed later. An optimized ML network is developed by stacking convLSTM layers vertically. The ML architecture is then compiled and trained using the input samples. The trained models are saved as .h5 files for deployment. These files are accessed later for predicted training frames (seen to ML), test frames (unseen to ML), and performance analysis. The sample preparation for the test set is different from the training sample. Starting with the last few frames from the training set, the predicted frames are added back to the sample as part of input frames as shown in the workflow (Figure 2) by an arrow line indicating this action. For performance analysis, the execution times for training, predicting training frames, predicting test frames are recorded. For quantitative analysis, normalized mean square error and structural similarity index measure (SSIM) between two frames are calculated (Appendix B). The difference frames for visual comparison are also generated.

**Machine Learning Model**

***Sample Preparation***

The extracted 2-D frames of properties are often not directly used as the inputs to machine learning algorithms. A specific machine learning algorithm requires a special sample preparation from the original data. All $P$ numbers of different dynamical properties (oil, gas, and water saturations and pressure) are required to monitor over the entire period of production and injection operations. These properties are recorded at all locations represented by an M × N grid (34 × 16). These spatial observations of P different properties at a single instant can be expressed by a tensor $X \in \mathbb{R}^{PXMXN}$. A sequence of these tensors ($X_1, X_2, X_3, \ldots X_T$) is formed for temporal representation for the entire period of operation. Using these tensors, suitable samples are prepared as inputs to convLSTM. In our convLSTM model, the number of output frames is the same as the input frames. The sample can be prepared in two ways, as shown in Figure 3.

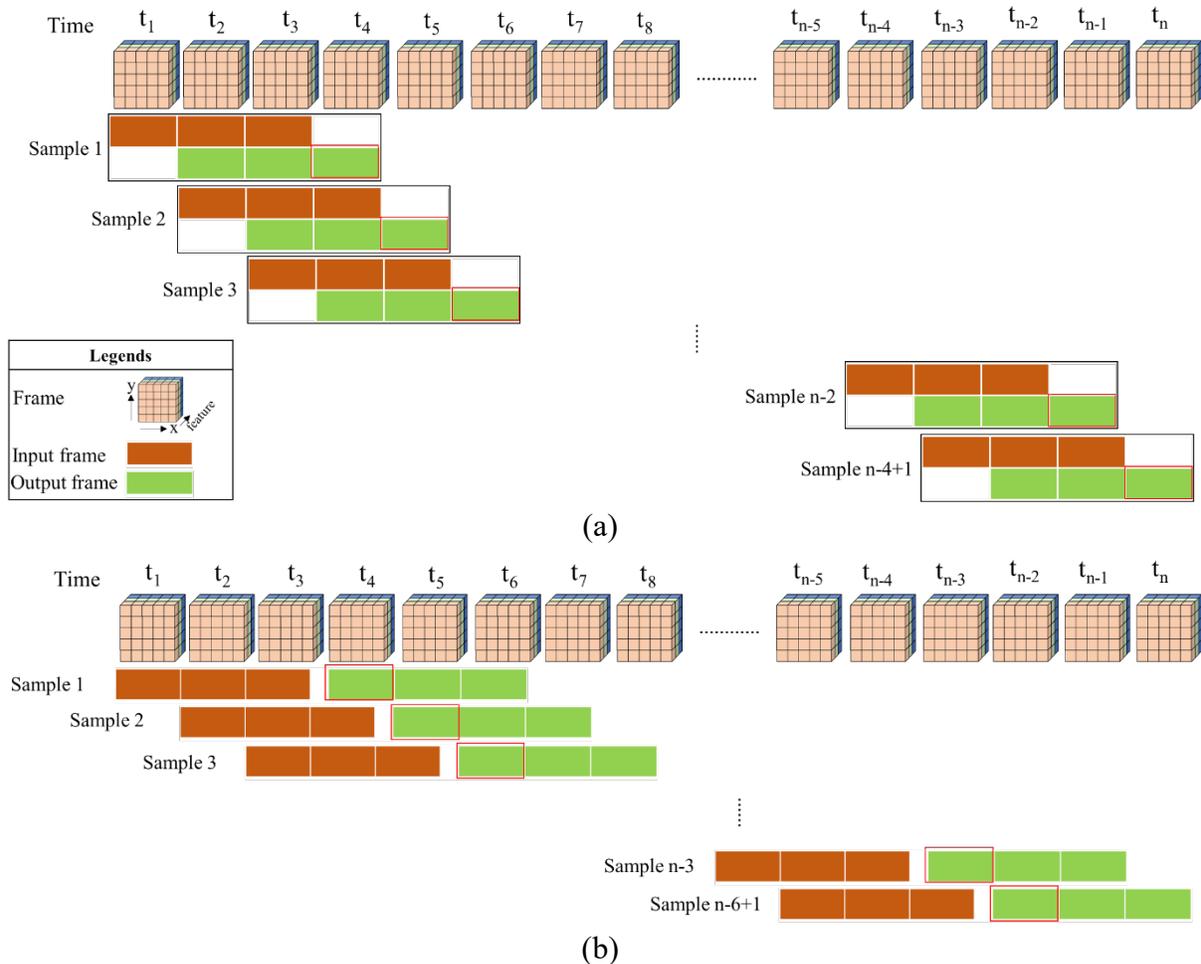

Figure 3: (a) samples consist of the overlapping input and output frames by shifting output frames by one time-step forward (b) samples with non-overlapping input and output frames

In the first method (or overlapping method) shown in Figure 3a, a number of sequential frames are taken as inputs (as marked by brown boxes), and the same number of frames forwarded by a single time step is taken as outputs (marked by green boxes). This entire set of input and output frames makes a single sample. During testing of the trained model, it is often desired to predict a single output frame for multi-step ahead predictions. In those instances (that is also the case for this study), the last predicted frame (that is not overlapped with input frames) as marked by the red line is considered as a single output.

In the second method or non-overlapping method, as shown in Figure 3b, a few sequential frames (marked by brown boxes) are taken as input, and the same number of next sequential frames (marked by green boxes) are considered as outputs. Unlike the overlapping method, the first predicted frame as marked by the red line is considered as a single output during testing. The selection of the number of frames for each sample is based on a priori knowledge and the amount of available training data. In this study, the overlapping method of sample generation is adopted based on our preliminary investigations. The overlapping method showed a better prediction compatred to non-overlapping method for our system. Though it is worth noting that the selection of the method of sample generation is at domain expert's discretion.

### *Network Structure and Implementation*

An optimum machine learning model architecture is developed to obtain the best possible results in a reasonable time using less computational memory. Spatial and temporal correlations are modeled using convolutional long short-term memory (convLSTM) in the core of the entire machine learning algorithm. After the initial investigations on the number of different ML layers such as convLSTM, batch normalizations, conv3D, an optimum structure is created. The first two layers are two-dimensional convLSTM, and the last two layers are three-dimensional convolutional models. Batch normalization is also applied before the first three layers to standardize the inputs for each batch of samples. This technique also stabilizes the learning process and reduces the number of epochs during training [32]. The architecture of this ML algorithm is shown in Figure 4**.**

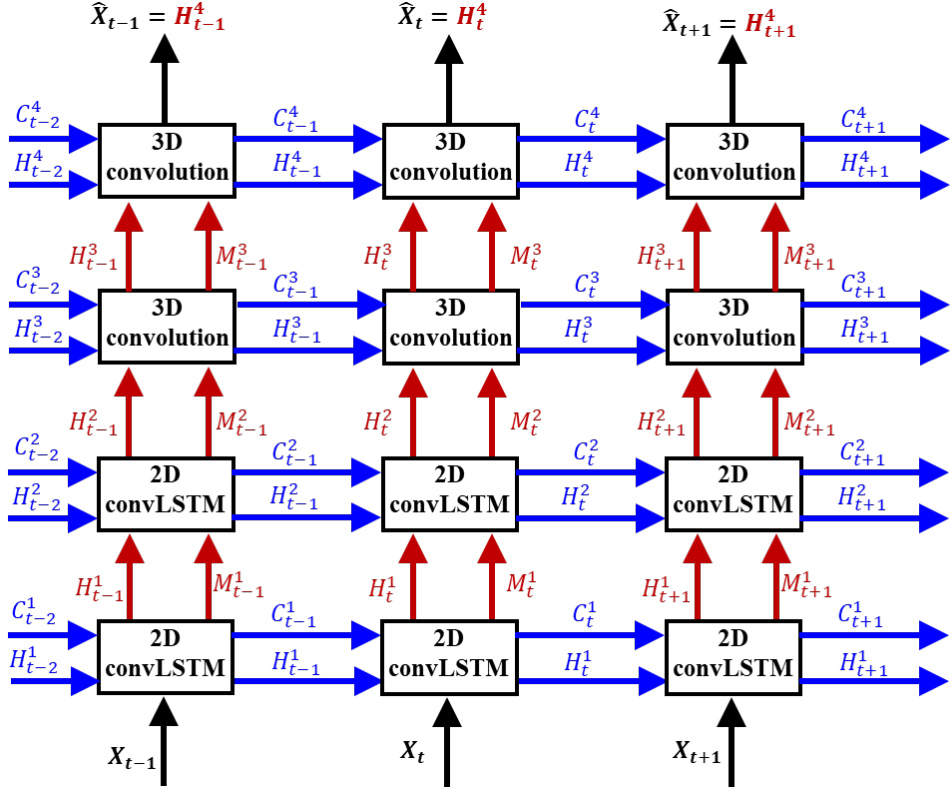

Figure 4: Machine learning architecture using convolutional long short-term memory (convLSTM): the brown lines indicate the deep transition paths of the spatial memory, while horizontal blue arrows indicate the update directions of the temporal memories.

A convLSTM unit represents spatial and temporal properties in a unified memory cell and transfers them vertically across layers and horizontally over time. Therefore, one vertical stack carries information for the same time step from bottom to top. The input frames of a sample (brown boxes in Figure 3) are fed into the first layer or bottom layer, and the output frames are predicted from the top layer. The spatial memory and hidden states flow vertically from bottom to top in the layers, and the temporal memory and hidden states flow horizontally in the next time step. The information flow and associated equations in a single unit of convLSTM are discussed in Appendix B.

The entire code is written on a Python notebook namely Spyder [33] which is an open-source, cross-platform integrated development environment (IDE) available on Anaconda Navigator [34]. All machine learning libraries (convLSTM, 3D convolution, etc.) and compiler with optimizer are imported from Keras [35] that is built on TensorFlow [36] for efficient numerical computations.

### Training & Testing

Pressure and oil, gas, and water saturation maps on the 21$^{st}$ horizontal layer at the end of each month were used for the machine learning algorithm. Therefore, 360 maps for each property were available from 30 years of operation in the SACROC unit. The data from the first 25 years (approximately 83%) are used for training the convLSTM model network, and the rest data from the last five years are kept aside for testing the trained model. A sample consists of 10 frames, i.e., frames #1 to #10 are taken as the input frames for the first sample and frames #2 to #11 as out frames for the same sample as described in details in the earlier section (Figure 3a). Similarly, frames #2 to #11 are the inputs for the second sample, and frames #3 to #12 are the outputs. It is noted that the starting frame in each sample is moved by a one-time step forward. Therefore, a total of 290 samples is generated from 300 frames considering ten frames per sample in the overlapping sample generation method. During training, the machine learning model predicts 290 frames as output, and they are compared with the actual frames to calculate the loss function. Getting back this feedback of loss function, the numerical optimizer in the algorithm then adjusts all weights and biases to minimize the loss function. The number of ML parameters in each layer for different trained models are shown in Appendix C. Strategically, it is not suggested to use the entire samples to train the model at once but use the samples in batches, and the learning from one batch is transferred to the next one. This strategy works fine for a larger data set. A batch size of 5 is used in this study.

The number of convolutional filters and their sizes are adjusted in each layer (layers 1 to 4 as shown in Figure 4) during training. This requires a significant understanding of the behavior of the network in response to the convolutional operation. The number of filters in each layer for different models is tabulated in Appendix C. Rectified linear unit (ReLU) and sigmoid activations function are used in the conv3D layers (3$^{rd}$ and 4$^{th}$ layers). The mean square error (MSE) is used as the loss function. The Nesterov accelerated adaptive moment estimation *(*nadam*)* optimizer is found to be effective for this network.

Unlike the prediction of training frames, sample preparation for the prediction of test frames works differently. The same set of samples that were assigned during training the models are used to predict the training frames as described in Figure 3. However, in the case of the multi-step ahead predictions of test frames, the last ten trained frames, i.e., 291 to 300, are used as the initial inputs to the trained model. This predicted frame is then assigned as one of the input frames along with the last nine frames to predict the next one. This method is also described in the workflow section earlier. Subsequently, after the ten predictions, the entire ten frames in the input consists of the predicted frames. Thereafter, all test frames are predicted from the previous predictions. Therefore, it is critically important to obtain good predictions for the first few frames. Otherwise, the errors in the predictions carry forward in the future.

During the training of different models of pressure, oil saturation, and gas saturation, the mean squared errors (MSE) sharply decline and then stabilize after a few epochs, as shown in Figure 5. A maximum of 30 epochs is chosen to avoid any unnecessary computational time once the solution is reached. It is also noted that a low number of epochs may result in an incomplete training of the models.

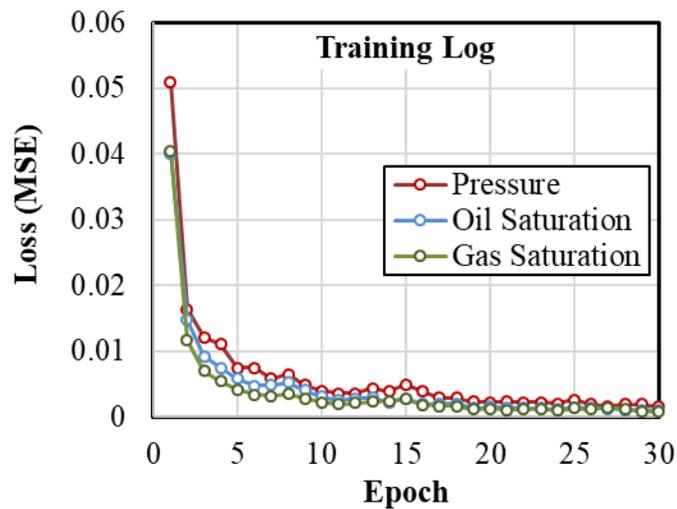

Figure 5: Minimization of loss function during training of convLSTM models for pressure, oil saturation, and gas saturation

The training of the ML model based on convLSTM, as shown in Figure 6, takes 8 to 13 minutes using a working laptop (Intel® Core™ i7-3630 QM, 2.40 GHz, 16.0 GB RAM). Trained models are saved for later to predict both the training frames (already seen by the models) and testing frames (unseen to the models). The execution times for 290 training frames and 60 test frames are recorded for all models. The average prediction time for training and testing is shown in Figure 6.

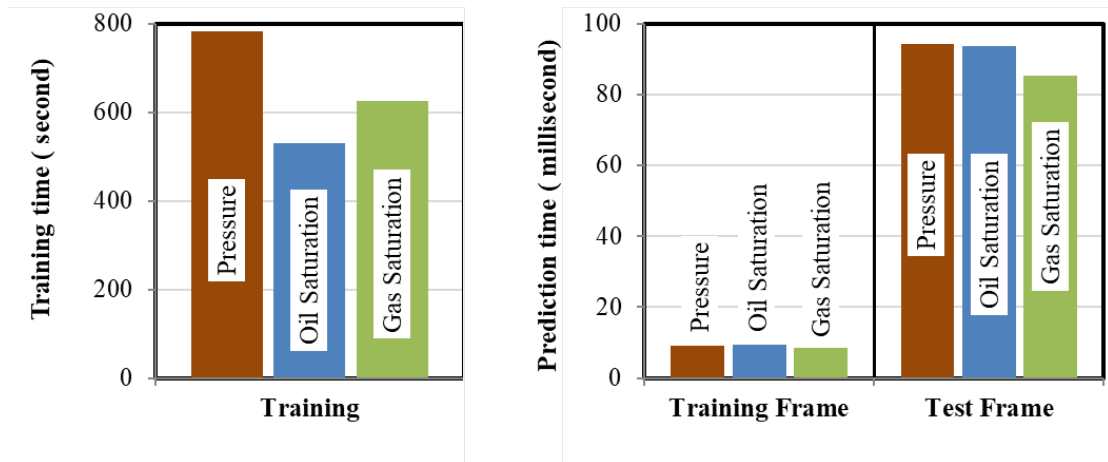

Figure 6: Execution times for training and average prediction time per frame of 290 training frames and 60 test frames

The prediction time per frame for train data is significantly lower (~9 ms) than the test prediction time (~90 ms). In both cases, the overall prediction time is practically negligible. The model took about 3 seconds to predict the first 25 years (training) of monthly pressure and saturation, whereas the total prediction time was about 6 seconds for the last five years' (test) monthly data. Because the input samples are prepared differently for test prediction as described earlier, prediction time is relatively higher. On the other hand, the input samples for training prediction are already prepared during training the ML model. Therefore, the computational time is less for predicting of training data set.

**Results and Discussions**

The predicted maps and actual maps of pressure, oil saturation and gas saturation are compared side by side for the visual comparisons along with the difference maps. For quantitative comparisons, structural similarity index measure (SSIM) and normalized root mean squared error (NRMSE) are calculated.

The acceptability of the ML models is predominantly evaluated by the accuracy in the predictions of the unseen test set. However, training and test sets are not assessed independently because the successful development of ML models depends on the training set. Often overfitting the models by closely matching all frames in the training set causes poor prediction or higher errors for blind data set (e.g., test set). Similar issues are observed for the underfitting of the models. Hence, both the results for training and test sets are discussed separately. There are 291 predicted frames for the training period (10th month to 300th month) and 60 predicted frames for the test period (301st month to 360th month). All frames are not shown in this article for the sake of brevity. Instead, only the first three frames and the last three frames are discussed for all three properties to assess the predictability of the ML model. However, SSIM and NRMSE are plotted for the entire period of training and test sets.

The results from the training period are provided in **Appendix A**. As shown in Figures A.1 to A.3, the predicted frames for pressure, oil saturation, and gas saturation during training closely match with the actual frames because of the nature of the tuning of the model. Consequently, a higher degree of accuracy is achieved, as shown in Figures 8, 10, and 12. For the pressure prediction for the training set, the SSIMs stay close to 1 (0.9 to 1) except first few months. Similarly, the NRMSEs are low between 4 to 6%. Periodic variations in both SSIM and NRMSE are observed due to the alternating injection cycles of gas and water, which are not properly captured by the ML model. For oil and gas saturation, SSIMs reach unity after a few months, and the NRMSEs are below 5%. Based on these SSIM and NRMSE values, the training of three ML models for pressure, oil saturation, and gas saturation is satisfactory, and hence they can be adopted as the properly trained models for the prediction of the blind test set.

The first predicted frame in the test period is theoretically to be better than the frames that are predicted in the later periods. This is the consequence of the way the inputs are prepared for the test prediction. The last ten ground-truth frames of the training set (i.e., 291st month to 300th months) are used as the inputs to the trained models to predict the first test frame, i.e., frame at 301st month. Then, the first predicted test frame and last nine ground-truth frames of the training set (i.e., 292nd month to 300th months) are used as inputs to predict the second test frame, i.e., frame at 302nd month. In this process, the 11th predicted frame, i.e., frame at 311th month ( and onwards) is entirely based on the predicted input frames. The procedure is continued until all frames for the entire test period of 60 months, i.e., 301st month to 360th month, are covered. These results are reflected in the difference maps and the plots of SSIM and NRMSE. Like training predictions, only the first three frames (301st to 303rd months) and the last three frames (358th to 360th months) are shown here for side-by-side comparison and difference frames. Pressure maps for the test period are shown in Figure 7.

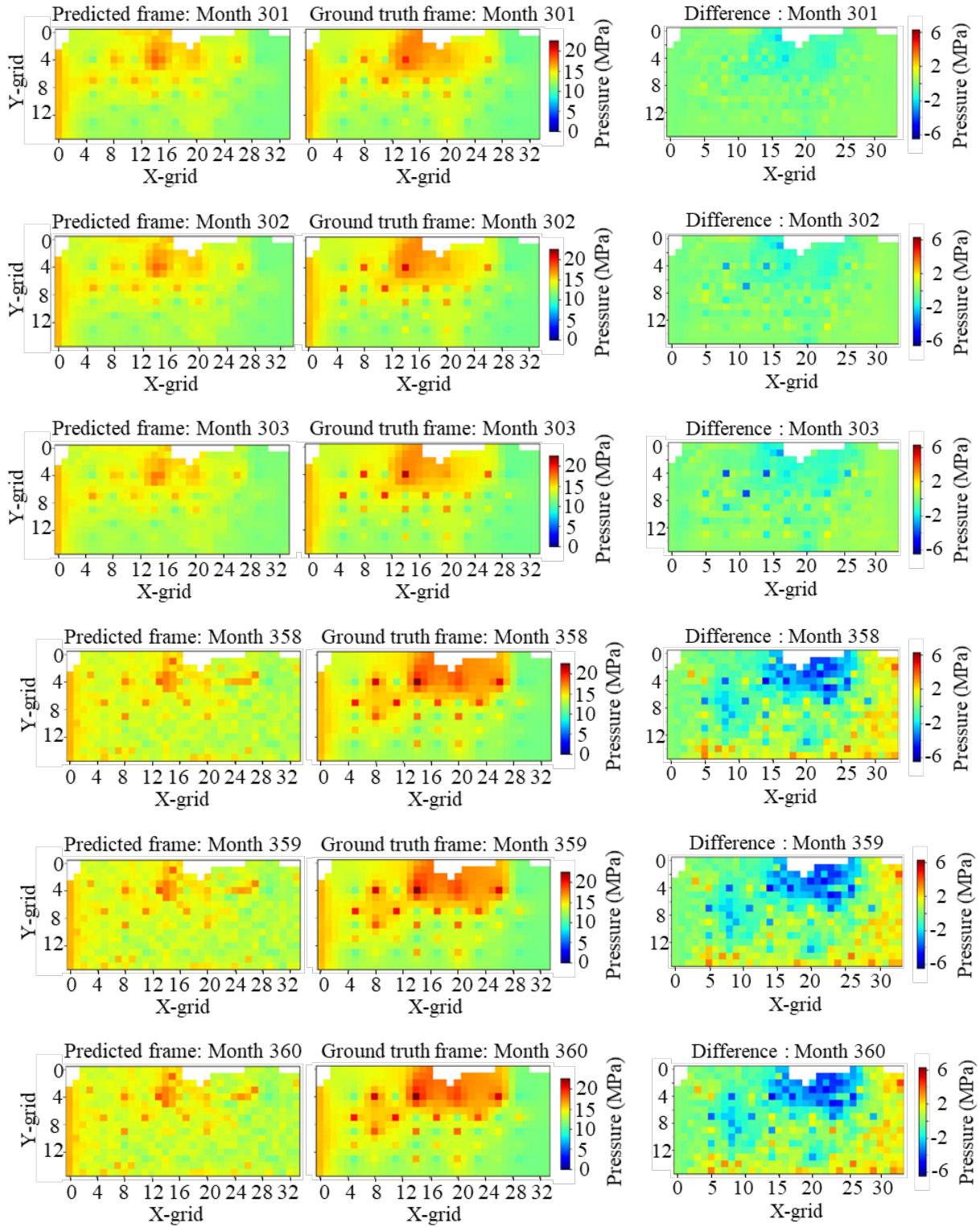

Figure 7: Frame to frame comparison of the first three and the last three pressure predictions of test data; the first column for predicted frames, the second column for ground truth frames, the third column for the difference in the predicted and ground truth frames

As explained before, the differences between the predicted frames (left column) and the ground-truth frames (middle column) are less in the first three frames compared to the last three frames. However, the higher intensity zones are rightly captured by the ML models. Additionally, these pressure differences are higher near the well locations. The possible reason is the drastic change in pressure in the injection wells during the cyclic injection of water and $CO_2$ and the operation changes in the production wells. The error analysis for train and test sets is shown in Figure 8.

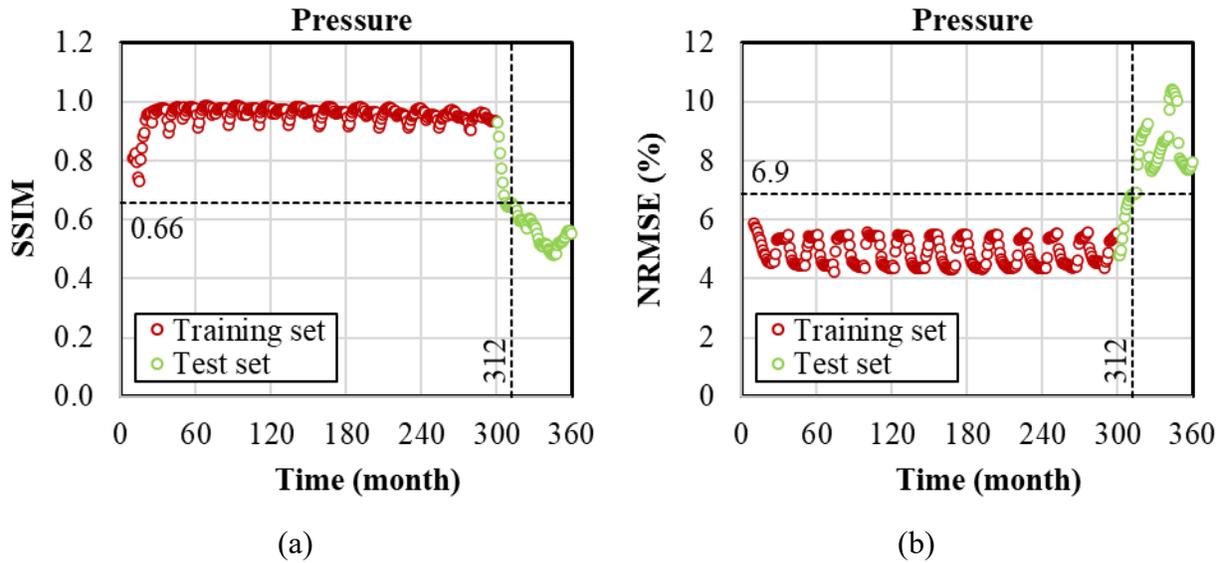

(a)                                                                    (b)

Figure 8: Error analysis of predicted pressure frames for training and test sets (a) Structural Similarity Index Measure (SSIM) (b) Normalized Root Mean Squared Error (NRMSE)

The vertical lines (dotted) in figure 8(a) and (b) mark the one year or 12 months ( i.e., from 301 month to 312 month)  in the testing time. The SSIM progressively drops from unity at the end of the training period to 0.5 at the end of the test period. Similarly, NRMSE increases from 5% to 9% in the same period. However, it is evident that the SSIM is greater than 0.66 and NRMSE is lower than 6.9% in the first 12 months of the test period. This error analysis confirms that the more dissimilarities between ground-truth and predicted frames in the later period of the test set. These disparities are anticipated, as described earlier. The predictions in the later stages are dependent on the previous predictions and therefore the initial predictions are vital. The differences between ground truth and the predictions in the first few months are carried forward cumulatively as time moves forward. Despite the differences, the errors are within the acceptance range in the first few months of blind predictions.

The blind predictions of the first three frames and the last three frames of oil saturation are shown in Figure 9.

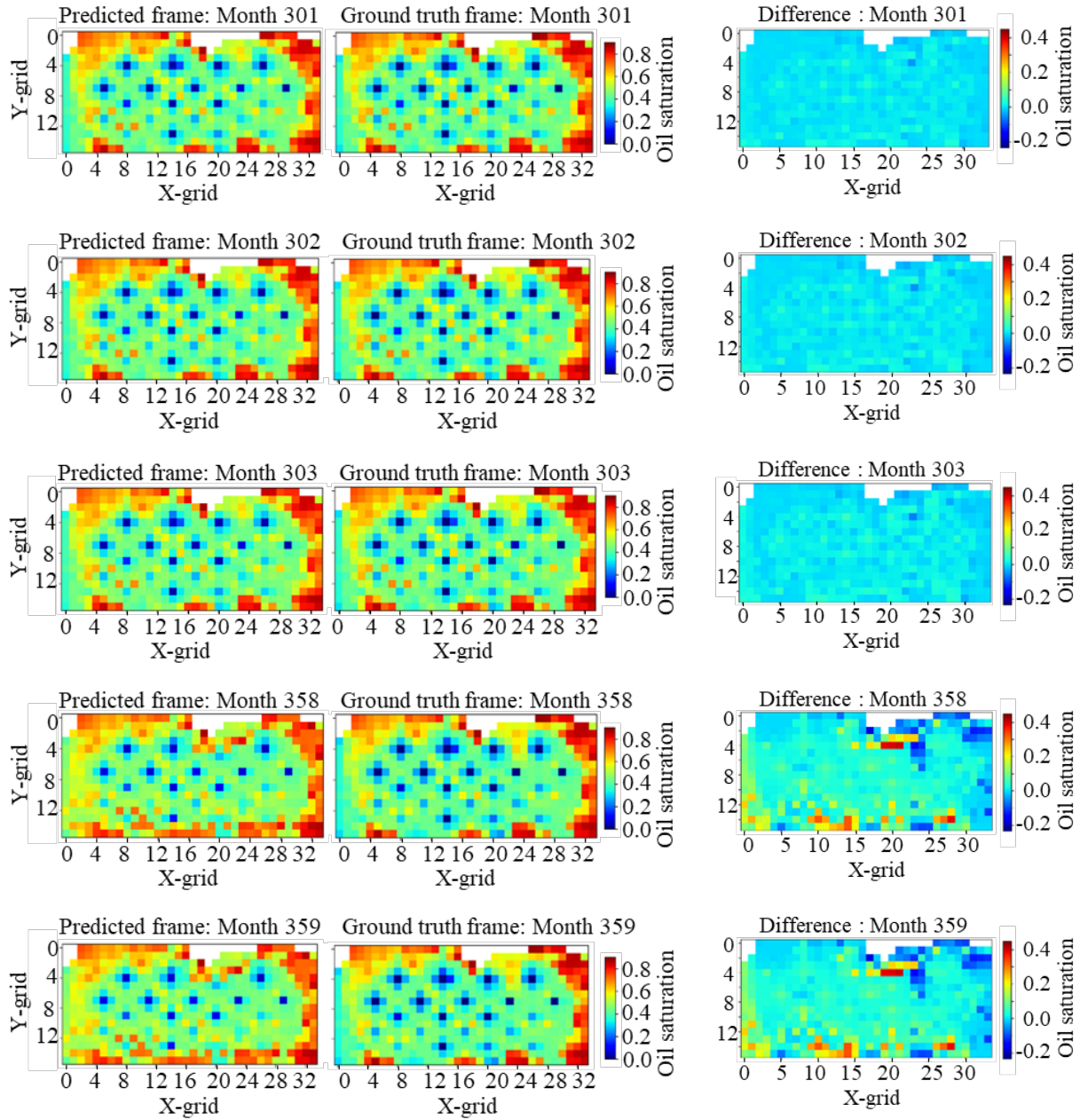

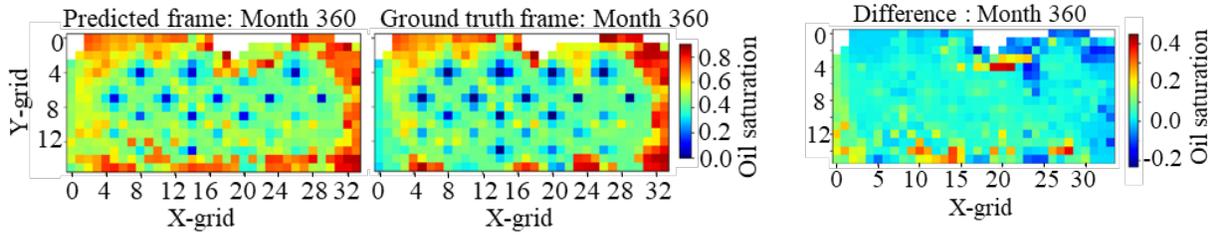

Figure 9: Frame to frame comparison of the first three and the last three oil saturation predictions of test data; the first column for predicted frames, the second column for ground truth frames, the third column for the difference in the predicted and ground truth frames

After the initial exploitation of hydrocarbon from the reservoir (mainly from the inner part of the reservoir), the oil rate reaches a steady value varying periodically with the injection cycle [27]. This happens towards the end of the training period, as seen from Figure A.2. The same trend continues in the test period. Therefore, the variation in oil saturation is low with time in the later stage of the production. In addition, some oil remains stagnant around the boundaries of the SACROC reservoir model for the entire period of 30 years. This causes higher oil saturations at these locations. The ML model has successfully captured these characteristics of the oil saturation. The first three predicted oil saturation maps are quite the same as the corresponding ground-truth maps. The difference in the oil saturation is only about 0.1. The last three predicted frames have a higher localized difference up to 0.4 in some areas. The differences in the oil saturation in the majority of the cells are only about 0.2. The quantitative error analysis of oil saturation is shown in Figure 10.

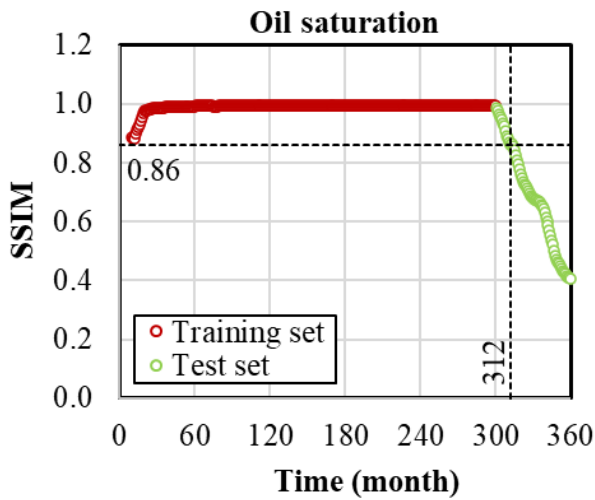

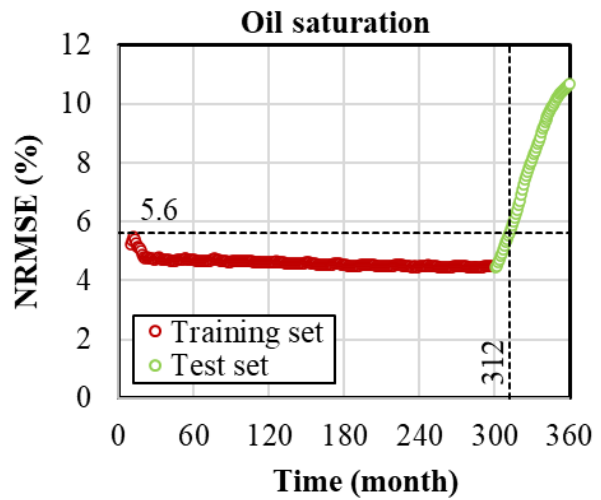

(a)                    (b)

Figure 10: Error analysis of predicted oil saturation frames for training and test sets (a) Structural Similarity Index Measure (SSIM) (b) Normalized Root Mean Squared Error (NRMSE)

Despite a good qualitative match and high accuracy in the first 12 months of test period(SSIM>0.86 and NRMSE < 5.6%), the SSIM steadily drops to 0.4 from 1.0. This is mainly due to the higher localized errors than the overall similarity in the trends of oil saturation. The NRMSE also increases to about 11% at the end of the test period.

The first three frames and the last three frames of gas saturation during the test period are shown in Figure 11.

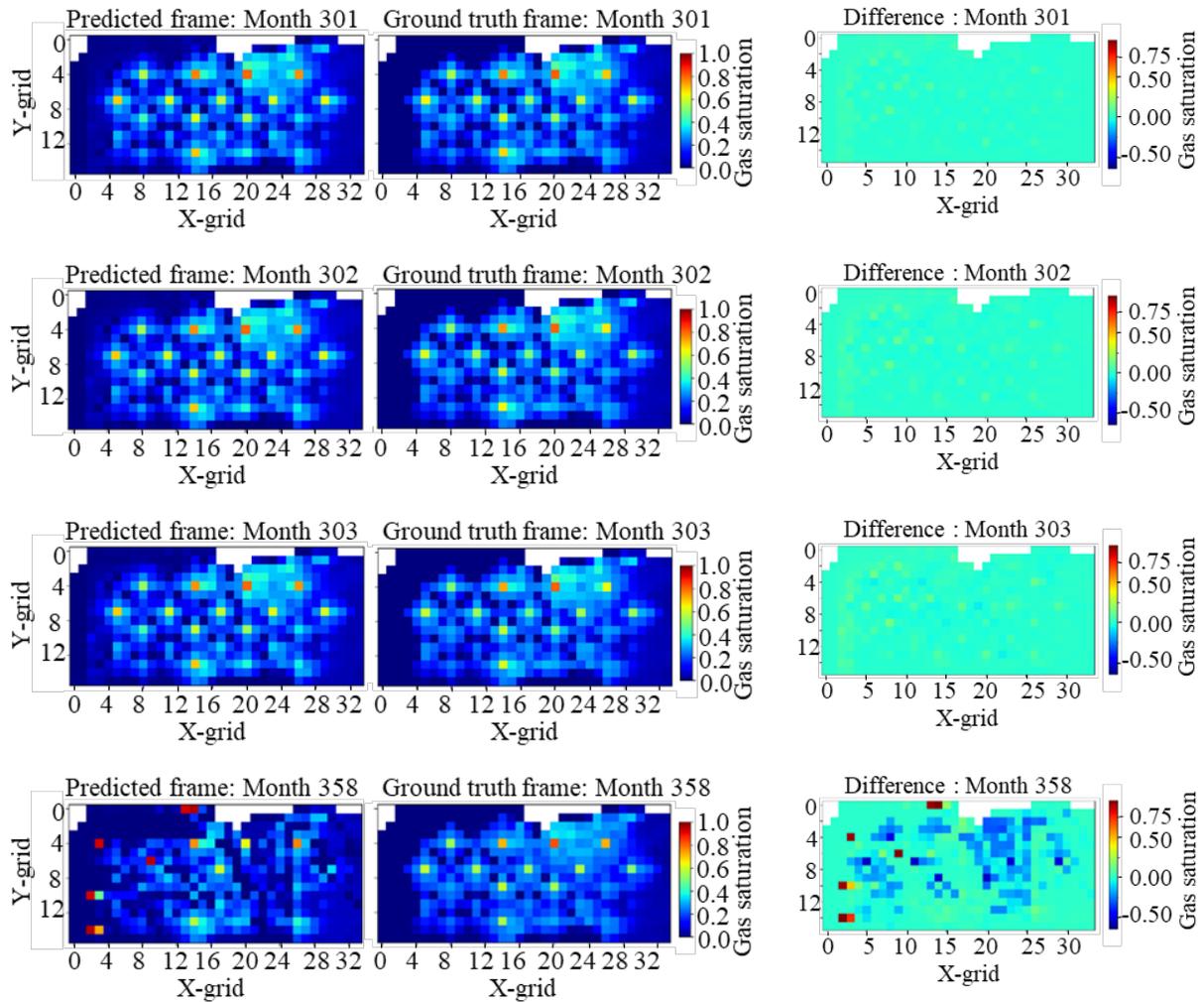

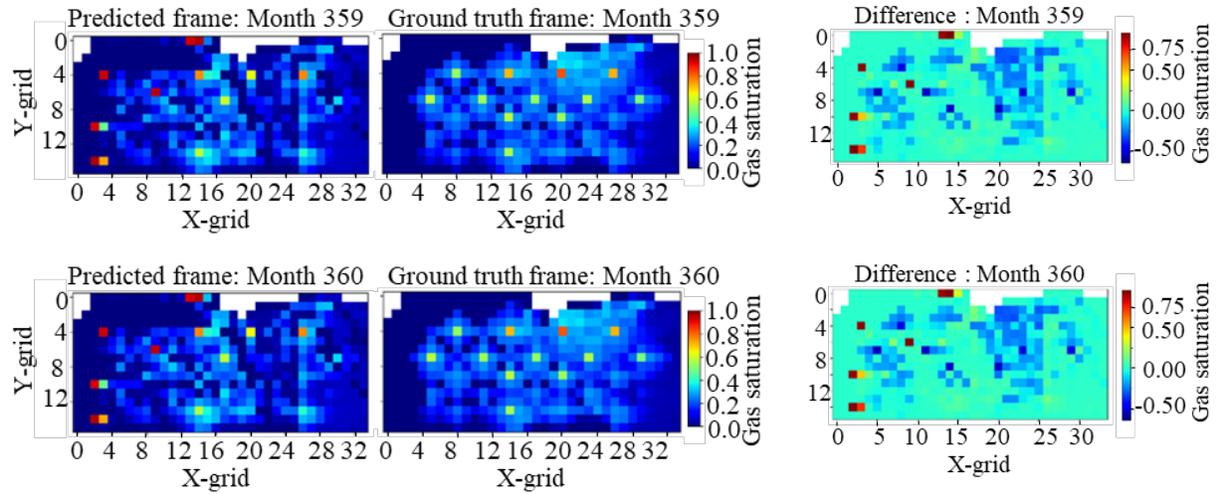

Figure 11: Frame to frame comparison of the first three and the last three gas saturation predictions of test data; the first column for predicted frames, the second column for ground truth frames, the third column for the difference in the predicted and ground truth frames

The SACROC reservoir model in this study is predominantly occupied by oil with less than 30% gas as seen in the first three frames of Figure A.3. The saturation especially around the producer wells decreases with time due to the production of gas. However, the cyclic injection of CO2 increases the gas saturation again near the injection wells. Therefore, sharp contrasts can be observed in the near well-bore regions. Except for these localized regions, the ML model successfully predicts the gas saturation in the other cells. As seen in the difference maps ( the third column in the figure), the difference in gas saturation is near zero other than the near well-bore regions. The SSIM and NRMSE for overall errors in the prediction of gas saturation are shown in Figure 12.

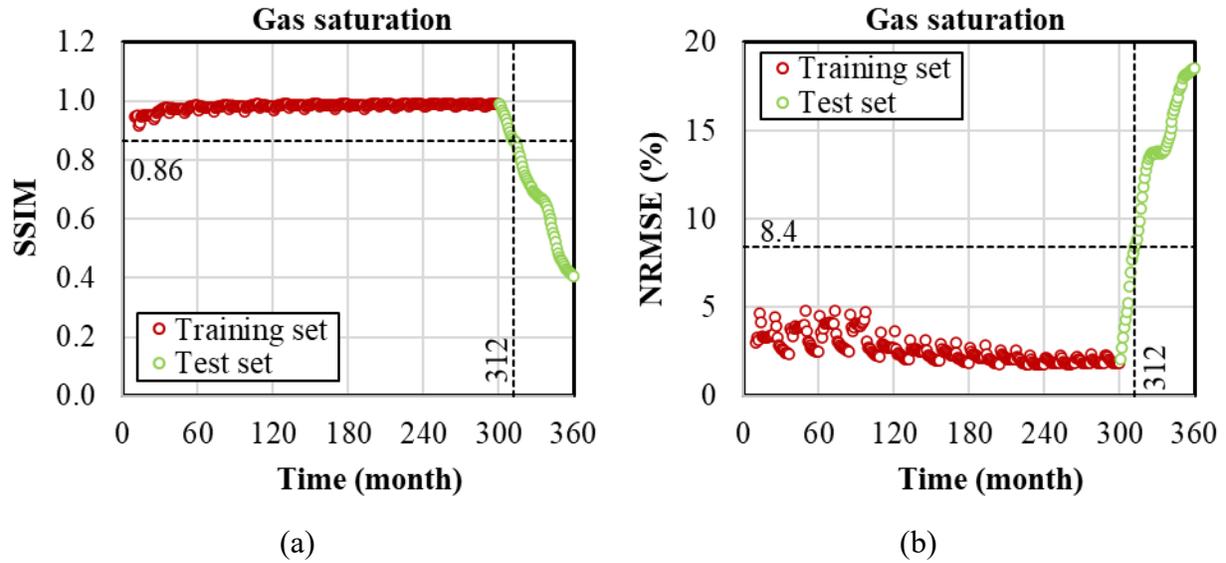

(a)                                                    (b)

Figure 12: Error analysis of predicted gas saturation frames for training and test sets (a) Structural Similarity Index Measure (SSIM) (b) Normalized Root Mean Squared Error (NRMSE)

Similar to oil saturation, the SSIM declines from unity to 0.4 at the end of the test period. The NRMSE also reaches 18% at the same time. The probable reason for the higher overall errors is the differences in the gas saturation in the well locations. Unlike oil saturation, the injection of $CO_2$ leads to an unusual variation in gas saturation.

**Conclusions**

The SACROC reservoir is a complex unit due to heterogeneities in permeability and porosity, well completion and injection and production operations. In this reservoir, the convolutional long short-term memory (convLSTM) is a suitable ML model to describe the flow in terms of pressure and saturation in the subsurface. The trained models using 25 years of monthly data of pressure, oil saturation and gas saturation achieved acceptable accuracy (SSIM>0.4, NRMSE < 20%) in predicting the future monthly maps for 5 years. The accuracy is even higher for the first few months. Therefore, it is demonstrated that the convolution LSTM is an efficient ML algorithm to capture the subsurface flow for spatio-temporal changes.

The workflow developed in this study for preparing samples and training of ML model is helpful in applying to any flow systems regardless of its complexity. The ML architecture can easily be modified by adding more layers and adjusting the number of neurons. These models are useful as precursors in the development of an integrated workflow in a data-driven model to understand the fluid flow in porous media.

## Funding

This work was performed in support of the National Energy Technology Laboratory's ongoing research under the RSS contract 89243318CFE0000003. The authors also acknowledge the academic license of CMG products from Computer Modeling Group, Calgary, Canada.

## Appendix A: Training Results

The first three and the last three predicted frames for the training set are compared side by side. The difference in the predicted frame and ground truth frame is also shown. Because the input samples are prepared from 10 frames (including the initial condition i.e., starting of the month), the first predicted frame is for the 10th month (the first input sample contains 0 to 9th-month frames). The last three predicted frames are for 298, 299, and 300th months.

The side-by-side comparisons and differences for the prediction of the pressure of the training set are shown in **Figure A.1**.

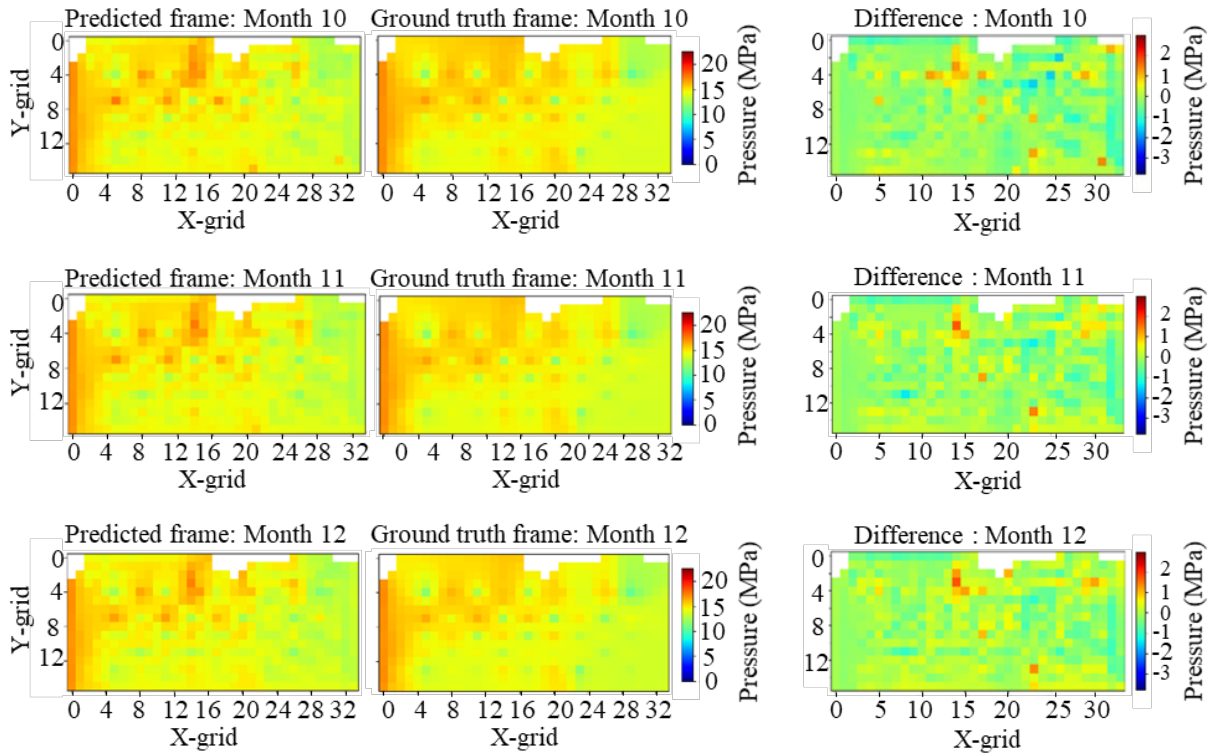

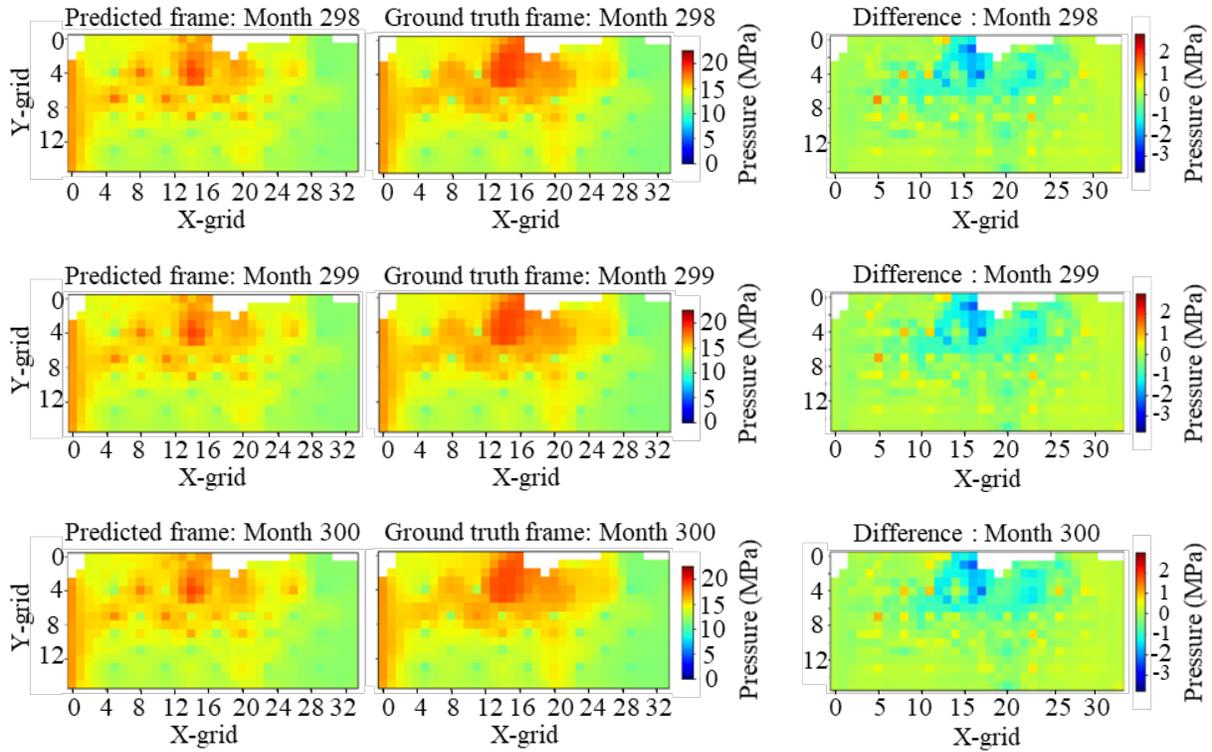

Figure A.1: Frame to frame comparison of the first three and the last three pressure predictions of training data; the first column for predicted frames, the second column for ground truth frames, the third column for the difference in the predicted and ground truth frames

The side-by-side comparisons and differences for the prediction of the oil saturation of the training set are shown in **Figure A.2**.

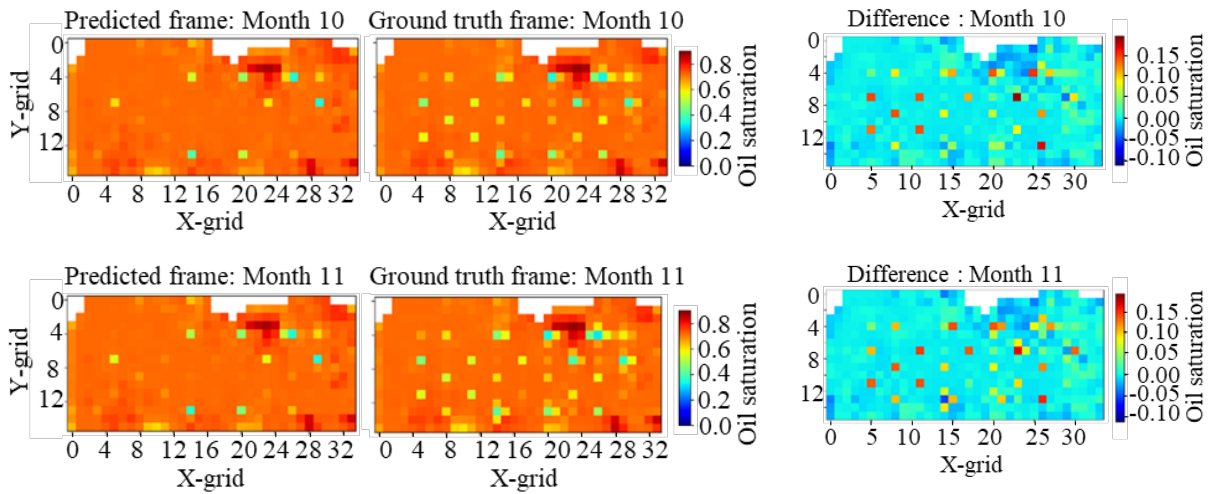

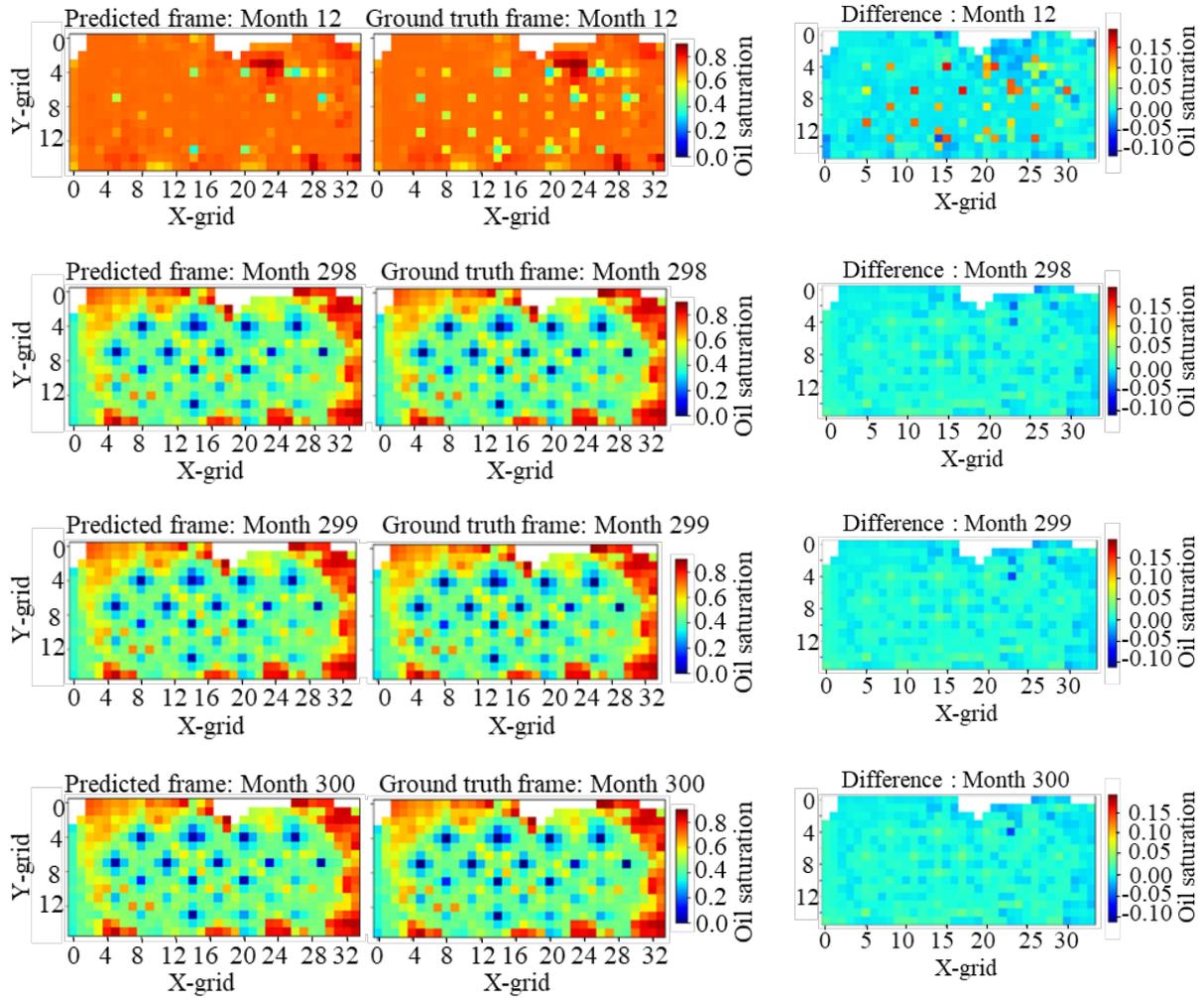

Figure A.2: Frame to frame comparison of the first three and the last three oil saturation predictions of training data; the first column for predicted frames, the second column for ground truth frames, the third column for the difference in the predicted and ground truth frames

The side-by-side comparisons and differences for the prediction of the gas saturation of the training set are shown in **Figure A.3**.

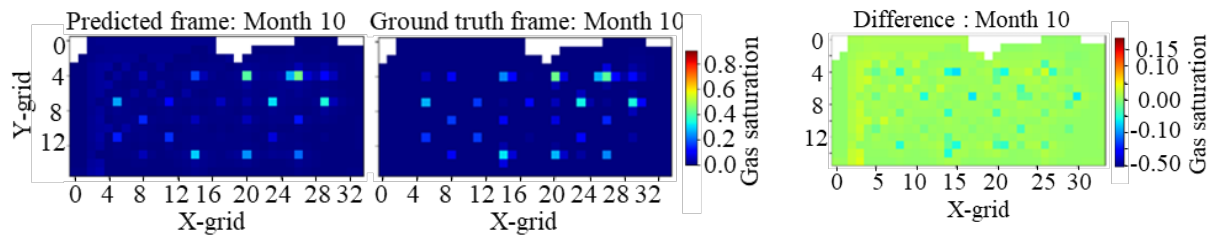

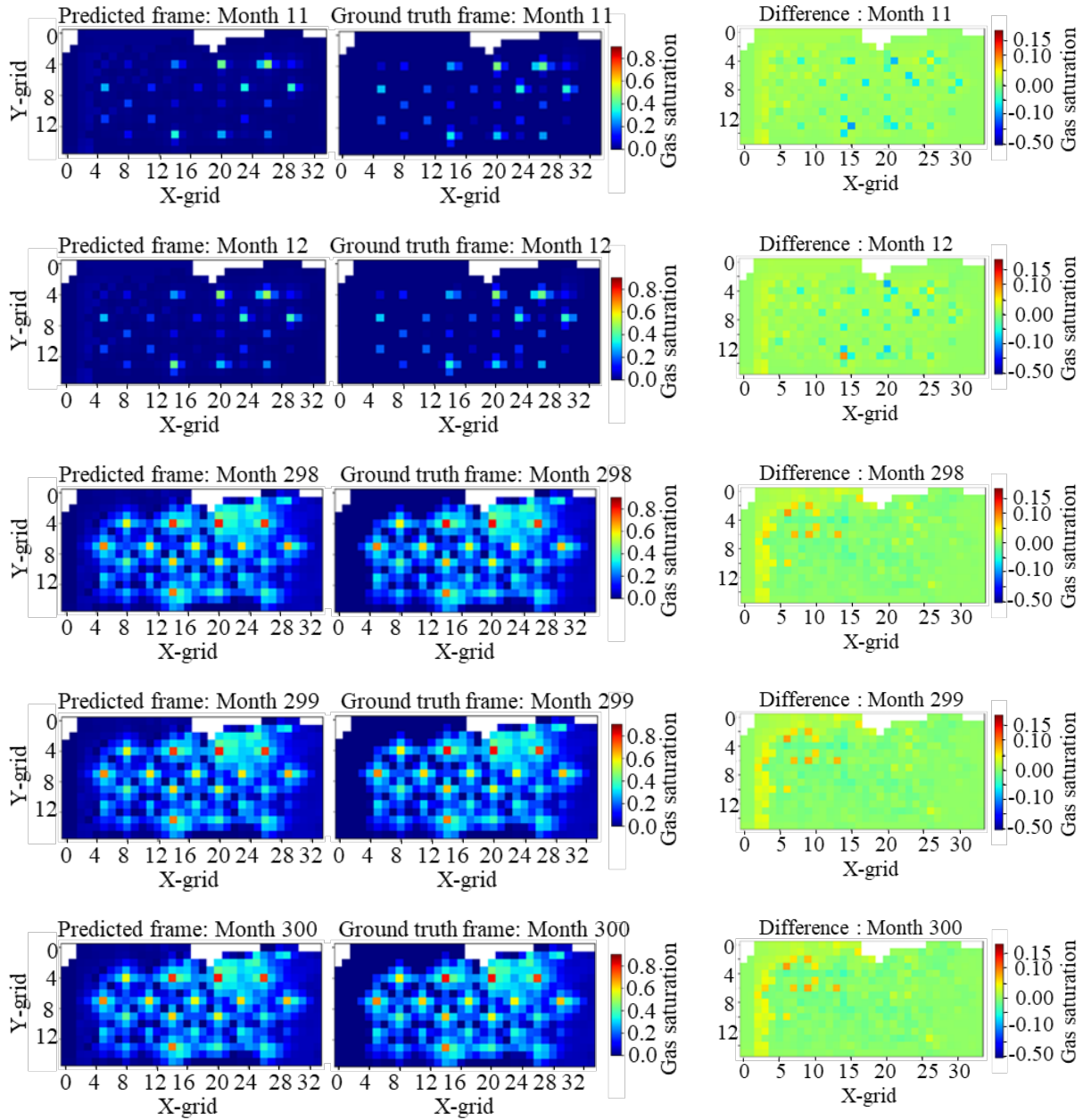

Figure A.3: Frame to frame comparison of the first three and the last three gas saturation predictions of training data; the first column for predicted frames, the second column for ground truth frames, the third column for the difference in the predicted and ground truth frames

## Appendix B: Error Analysis

Mean squared error (MSE) is used as a loss function during training of the machine learning (ML) model. MSE is calculated using **equation B.1.**

$$MSE = \frac{\sum_{i=1}^{n}(Y_{obs,i} - Y_{model,i})^2}{n} \qquad \ldots\ldots\ldots\ldots\ldots\ldots(B.1)$$

$Y_{obs}$ and $Y_{model}$ are the ground truth and predicted value from the ML model, respectively; n is the number of data points i.e., number of active grids which is 522 on the horizontal layer 21 of the coarse grid SACROC reservoir model in this study. Root mean squared error (RMSE) which is simply the square root of MSE is also frequently used to measure the overall error.

During training, all data sets are normalized data between 0 to 1. However, to assess the accuracy of the predicted results, data are transformed back to the actual values before calculating the errors. Pressure, oil saturation and gas saturation have different ranges of values (especially pressure and saturations). Therefore, MSE or RMSE varies greatly for the chosen parameters. It is often suggested to use a normalized root mean squared error (NRMSE) to compare errors for different parameters having wide ranges of data. The calculation of percentage NRMSE is shown in **equation B.2**.

$$NRMSE = \frac{\sqrt{MSE}}{Y_{obs,max} - Y_{obs,min}} \times 100 \qquad \ldots\ldots\ldots\ldots\ldots(B.2)$$

Where, $Y_{obs,\,max}$ and $Y_{obs,\,min}$ are the maximum and minimum values in the ground truth data set respectively.

The Structural Similarity Index (SSIM) is a perceptual metric that is used to compare two images: original image and degraded image (due to processing). Researchers [37-42] showed that the SSIM is a significantly better metric compared to MSE for the perceptual comparison of two images. The SSIM is defined in **equation B.3**.

$$SSIM\,(x,y) = \frac{(2\mu_x\mu_y + c_1)(2\sigma_{xy} + c_2)}{(\mu_x^2 + \mu_y^2 + c_1)(\sigma_x^2 + \sigma_y^2 + c_2)} \qquad \ldots\ldots\ldots\ldots\ldots(B.3)$$

where,

$$\sigma_{xy} = \frac{1}{N-1}\sum_{i=1}^{N}(x_i - \mu_x)(y_i - \mu_y) \qquad (B.4)$$

$$\mu_x = \frac{1}{N}\sum_{i=1}^{N}x_i \qquad (B.5)$$

$$\mu_y = \frac{1}{N}\sum_{i=1}^{N}y_i \qquad (B.6)$$

$$\sigma_x = \sqrt{\frac{1}{N-1} \sum_{i=1}^{N} (x_i - \mu_x)^2} \tag{B.7}$$

$$\sigma_y = \sqrt{\frac{1}{N-1} \sum_{i=1}^{N} (y_i - \mu_y)^2} \tag{B.8}$$

The value of SSIM close to unity shows a good match between an original image and the processed image.

## Appendix C:  Machine Learning Programs and Parameters

Four main scripts are written in this study for different purposes namely data extraction from output file generated by a reservoir simulator, preprocessing, ML model and training, and post-processing. The information flow and the connectivity among python programs are shown in **Figure C.1**.

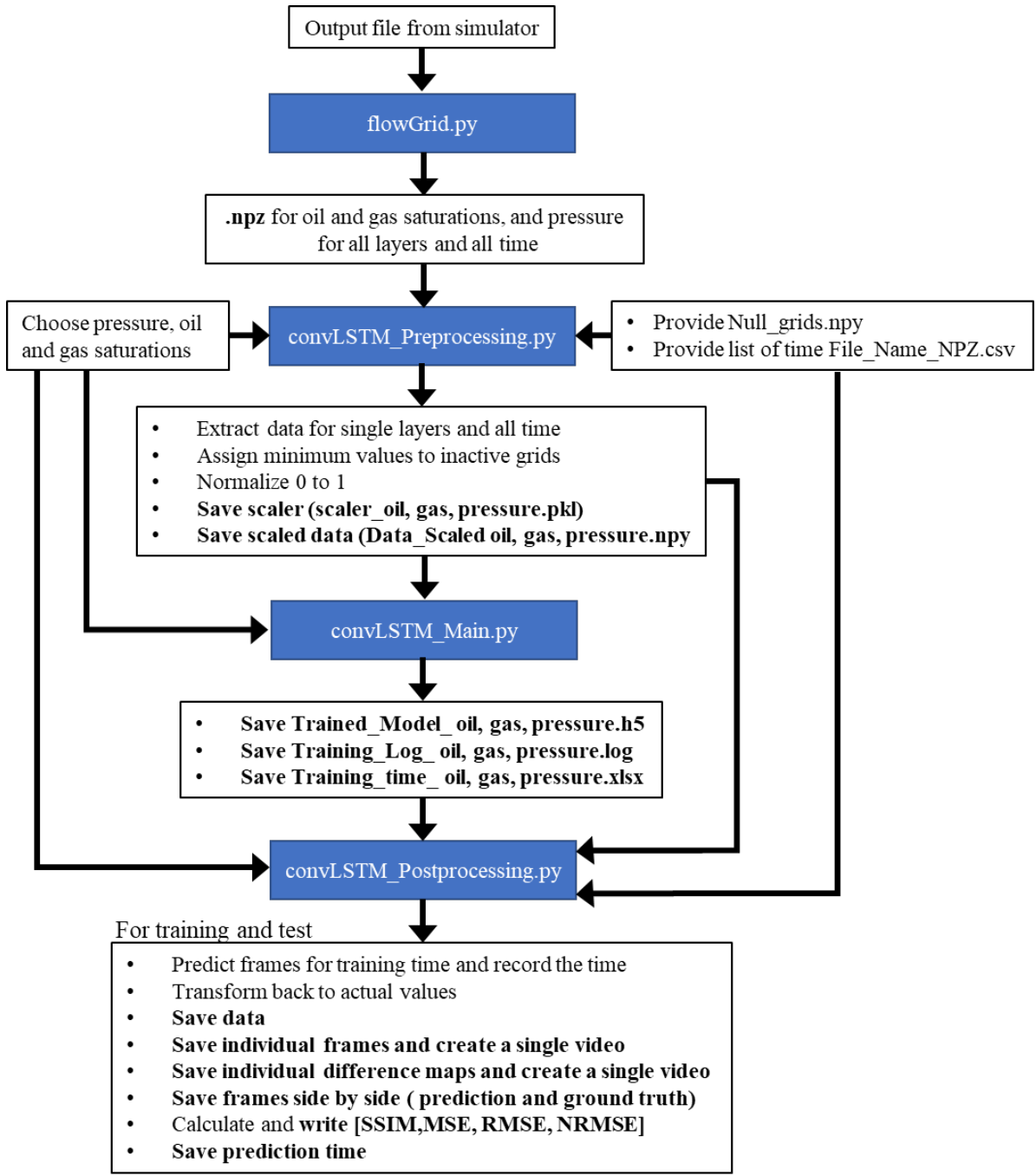

Figure C.1: Flow of information and the connectivity among scripts

The various procedures in this workflow are discussed in the 'workflow' section.

As shown in Figure 4, there are two convLSTM and two 3D convolutional layers. Each layer consists of multiple convolutional filters of different sizes. Sixteen filters of size 3x3 are used in the first convLSTM layer, 8 filters of size 3x3 in the second convLSTM layer and 4 filters of size 3x3x3 in the 3D convolutional layer and one filter of size 1x1x1 in the final 3D convolutional layer. The three batch normalization layers are also added before the first three layers. The same ML parameters and network work well for all three properties (pressure, oil saturation and gas saturation). The number of parameters in each layer of ML model is shown in Table C.1.

Table C.1: Number of ML parameters in each layer

| ML layer | Number of parameters |
|---|---|
| convLSTM 1 | 9856 |
| convLSTM 2 | 6944 |
| 3D convolution 1 | 868 |
| 3D convolution 2 | 5 |
| All batch normalizations | 100 |
| **Total** | **17773** |

A total of 17723 parameters (out of 17773) of the ML model is trainable that are obtained during the training process.

## Appendix D: Brief theory of convLSTM

Convolutional Long Short-Term Memory (convLSTM) for spatio-temporal prediction is briefly discussed here. A basic structure of the flow of information in convLSTM is shown in Figure D.1.

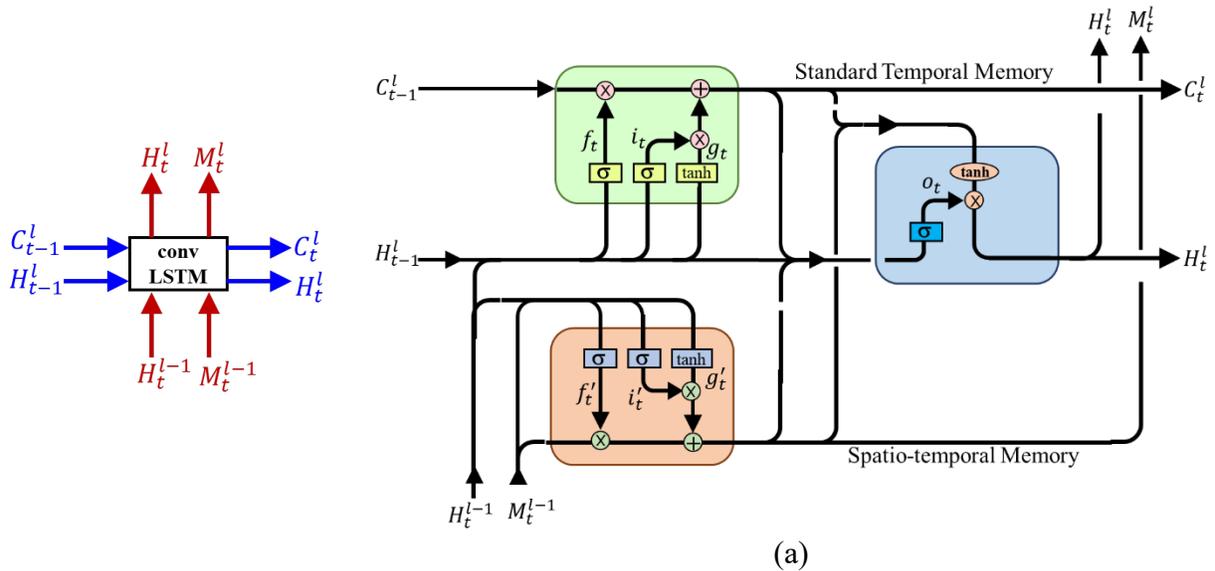

(a)

**Figure D.1**: convLSTM (ST-LSTM) cell used (modified from [3])

All the outputs i.e., h from each time step are compared with the actual values by calculating the mean squared error or other loss function. Finally, this loss function is minimized during the optimization of all the weights (W), and bias terms (b).

The classical LSTM only represents temporal memory whereas in the convLSTM model both spatio and temporal memory are represented. A single convLSTM model contains two input gates and two forget gates (one for temporal and one for spatial) and one output gate combining these two memories. The purpose of these gates is to control the flow of information. Various mathematical operators and notations used in Figure B.1 are shown in **equations D.1 to D.10.**

$$g_t = \tanh(W_{xg} * X_t + W_{hg} * H_{t-1}^l + b_g) \tag{D.1}$$

$$i_t = \sigma(W_{xi} * X_t + W_{hi} * H_{t-1}^l + b_i) \tag{D.2}$$

$$f_t = \sigma(W_{xf} * X_t + W_{hf} * H_{t-1}^l + b_f) \tag{D.3}$$

$$C_t^l = f_t \otimes C_{t-1}^l + i_t \otimes g_t \tag{D.4}$$

$$g_t' = tanh(W_{xg}' * X_t + W_{mg} * M_t^{l-1} + b_g') \tag{D.5}$$

$$i_t' = \sigma(W_{xi}' * X_t + W_{mi} * M_t^{l-1} + b_i') \tag{D.6}$$

$$f_t' = \sigma(W_{xf}' * X_t + W_{mf} * M_t^{l-1} + b_f') \tag{D.7}$$

$$M_t^l = f_t' \otimes M_t^{l-1} + i_t' \otimes g_t' \tag{D.8}$$

$$O_t = \sigma(W_{xo} * X_t + W_{ho} * H_{t-1}^l + W_{co} * C_t^l + W_{mo} * M_t^l + b_o) \tag{D.9}$$

$$H_t^l = O_t \otimes tanh(W_{1X1} * [C_t^l, M_t^l]) \tag{D.10}$$

**Nomenclature**

| Symbol | Description | Units |
|--------|-------------|-------|
| $\otimes$ | Pointwise Multiplication | - |
| b | Bias | Unit |
| convLSTM | Convolutional Long Short-Term Memory | |
| $C_t$ | Cell State | Unit |
| EDX | Energy Data Exchange | |
| EOR | Enhanced Oil Recovery | - |
| FSTN | Flexible Spatio-Temporal Network | |
| $f_t$ | Forget Gate | Unit |
| $g_t$ | One Output from Input Gate | Unit |
| h | Hidden Output | Unit |
| I | Injector Well | - |
| IDE | Integrated Development Environment | - |
| $i_t$ | Input State | Unit |
| LSTM | Long Short-Term Memory | - |
| ML | Machine Learning | - |
| MSE | Mean Squared Error | Square of Unit |
| NAdam | Nesterov Accelerated Adaptive Moment Estimation | - |
| NETL | National Energy Technology Laboratory | |
| NRMSE | Normalized Root Mean Squared Error | - |
| $O_t$ | Output from Ouput Gate | Unit |
| P | Producer Well | - |
| Physics-Informed ST-LSTM | Physics-Informed Spatio-Temporal Long Short Term Memory | |
| PredRNN | Predictive Recurrent Neural Network | |
| ReLU | Rectified Linear Unit | - |
| RMSE | Root Mean Squared Error | Unit |
| RNN | Recurrent Neural Network | - |
| SACROC | Scurry Area Canyon Reef Operators Committee | - |
| SSIM | Structural Similarity Index Measure | |
| SST | Sea Surface Temperature | |
| STA-LSTM | Spatio-Temporal Attention Long Short Term Memory Model | |
| Tanh | Hyperbolic Tangent Function | - |
| W | Weight | - |

## Declarations

### *Funding*

This work was funded by the Department of Energy, National Energy Technology Laboratory, an agency of the United States Government, through a support contract with Leidos Research Support Team (LRST). Neither the United States Government nor any agency thereof, nor any of their employees, nor LRST, nor any of their employees, makes any warranty, expressed or implied, or assumes any legal liability or responsibility for the accuracy, completeness, or usefulness of any information, apparatus, product, or process disclosed, or represents that its use would not infringe privately owned rights. Reference herein to any specific commercial product, process, or service by trade name, trademark, manufacturer, or otherwise, does not necessarily constitute or imply its endorsement, recommendation, or favoring by the United States Government or any agency thereof. The views and opinions of authors expressed herein do not necessarily state or reflect those of the United States Government or any agency thereof.